\PassOptionsToPackage{draft}{hyperref}

\documentclass[conference,letterpaper]{IEEEtran}
\ifCLASSINFOpdf
\else
\fi
\usepackage{pgf}
\usepackage{pgfkeys}
\usepackage{pgfmath}
\usepackage{import}
\usepackage{graphicx}
\usepackage{cite}
\usepackage{algorithm}
\usepackage[noend]{algpseudocode}
\usepackage{algorithmicx}
\usepackage{subcaption}
\usepackage{tabularx}
\usepackage{multirow}

\usepackage{amsmath}
\usepackage{amssymb}

\usepackage{url}
\begin{document}
% \bstctlcite{BSTcontrol}
%
% paper title
% Titles are generally capitalized except for words such as a, an, and, as,
% at, but, by, for, in, nor, of, on, or, the, to and up, which are usually
% not capitalized unless they are the first or last word of the title.
% Linebreaks \\ can be used within to get better formatting as desired.
% Do not put math or special symbols in the title.
\title{FALCON-C: Flow-based Analysis and Labeling for Connected Vehicular Network Cybersecurity
}
%
%
% author names and IEEE memberships
% note positions of commas and nonbreaking spaces ( ~ ) LaTeX will not break
% a structure at a ~ so this keeps an author's name from being broken across
% two lines.
% use \thanks{} to gain access to the first footnote area
% a separate \thanks must be used for each paragraph as LaTeX2e's \thanks
% was not built to handle multiple paragraphs
%

\author{Joshua~Bean and~Dimitrios~Michael~Manias \\ The Department of Computer Science and Engineering, Mississippi State University\\ jab1896@msstate.edu, dmanias@cse.msstate.edu}% <-this % stops a space

% note the % following the last \IEEEmembership and also \thanks - 
% these prevent an unwanted space from occurring between the last author name
% and the end of the author line. i.e., if you had this:
% 
% \author{....lastname \thanks{...} \thanks{...} }
%                     ^------------^------------^----Do not want these spaces!
%
% a space would be appended to the last name and could cause every name on that
% line to be shifted left slightly. This is one of those "LaTeX things". For
% instance, "\textbf{A} \textbf{B}" will typeset as "A B" not "AB". To get
% "AB" then you have to do: "\textbf{A}\textbf{B}"
% \thanks is no different in this regard, so shield the last } of each \thanks
% that ends a line with a % and do not let a space in before the next \thanks.
% Spaces after \IEEEmembership other than the last one are OK (and needed) as
% you are supposed to have spaces between the names. For what it is worth,
% this is a minor point as most people would not even notice if the said evil
% space somehow managed to creep in.

% The paper headers
\markboth{IEEE HPSR 2026}%
{Bean \MakeLowercase{\textit{et al.}}:Catchy title goes here}
% The only time the second header will appear is for the odd numbered pages
% after the title page when using the twoside option.
% 
% *** Note that you probably will NOT want to include the author's ***
% *** name in the headers of peer review papers.                   ***
% You can use \ifCLASSOPTIONpeerreview for conditional compilation here if
% you desire.

% If you want to put a publisher's ID mark on the page you can do it like
% this:
%\IEEEpubid{0000--0000/00\$00.00~\copyright~2015 IEEE}
% Remember, if you use this you must call \IEEEpubidadjcol in the second
% column for its text to clear the IEEEpubid mark.

% use for special paper notices
%\IEEEspecialpapernotice{(Invited Paper)}

% make the title area
\maketitle

\begin{abstract}
Along with the recent rise in popularity of Electric Vehicles (EVs), Electric Vehicle Supply Equipment (EVSE) has emerged as a new target for cyber attacks. Therefore, ensuring the security and integrity of network communication between EVSE components and vehicular clients is a significant challenge that must be addressed. To this end, this paper proposes a Flow-based Analysis and Labeling for COnnected vehicular Network Cybersecurity (FALCON-C) framework. The FALCON-C framework leverages an autoencoder for anomaly detection and is trained on a small number of benign flows from the CICEVSE2024 dataset. The model’s objective is to model benign flow behavior and identify malicious flows by detecting statistically different reconstruction error profiles. The results demonstrate that the model can successfully identify malicious flows, achieving 100\% accuracy. Initially, some benign flows were misclassified as malicious, resulting in a suboptimal false positive rate. A thorough analysis of the autoencoder's performance and the nature of misclassified flows led to the development of a refined decision boundary, improving the framework’s performance by 8.6 \%. FALCON-C is intended to support Security Operations Center activities by automating flow labeling, leading to the enhanced curation of reliable datasets that can be used for various activities, including threat modeling and hunting, decision auditing, and intrusion detection system refinement.
\end{abstract}


\maketitle

\section{Introduction}

Electric Vehicles (EVs) have quickly become mainstream, and, as with the rest of the Internet of Things (IoT) landscape, this has created new attack surfaces for hackers to exploit. Electric Vehicle Supply Equipment (EVSE) and, more broadly, EV charging infrastructure use various communication protocols, most notably the Open Charge Point Protocol (OCPP) and ISO 15118. These protocols enable communication among charging stations, the backend Charging Station Management Systems (CSMSs), and EVs.

With the constantly evolving threat landscape, the ability to access large volumes of annotated data for training Machine Learning (ML)-based Intrusion Detection Systems (IDSs) is becoming increasingly difficult. Manual annotation is oftentimes time-consuming, given the high volume of generated network traffic, and prone to human error. Aside from cybersecurity concerns, ML model lifecycle management is a critical priority in highly dynamic network environments to combat model drift and ensure performance reliability \cite{manias2023model}. To this end, developing an automated framework to correctly label large volumes of network data is critical for the longevity of ML models and for Security Operations Centers (SOCs) alike. This data will be used for various activities, including threat modeling, performance validation, and re-training. However, given the challenges of finding high-quality labeled data, a practical constraint of such a model is its ability to learn complex representations from a few labeled samples.

The autoencoder is a neural network architecture that falls under the category of semi-supervised learning. It implements a symmetrical architecture in which a set of input features is reduced to a latent space via hidden layers (encoding). After encoding, the latent features are expanded to the original feature set dimension via architecturally symmetrical hidden layers (decoding). Model performance is assessed by calculating the reconstruction error, that is, how closely the model outputs resemble the model inputs. This process of encoding and decoding has various benefits and practical applications, including dimensionality reduction and noise reduction. Most notably, these models excel at anomaly detection, where normal data is used to train the model, yielding a specific reconstruction error profile. In theory, anomalies would result in a different reconstruction error profile, which ideally can be identified and used to develop a threshold to distinguish between the two. 

To this end, this paper presents a \textbf{F}low-based \textbf{A}nalysis and \textbf{L}abeling for \textbf{CO}nnected vehicular \textbf{N}etwork \textbf{C}ybersecurity (FALCON-C). The FALCON-C framework leverages an autoencoder trained on the benign data in the CICEVSE2024 dataset \cite{Buedi2024} to generate automated ground-truth labels and perform post-flow high-confidence anomaly detection. The motivation behind this approach is that complete flows capture the full bi-directional communication patterns and behaviors of the system. This insight will assist with SOC activities, such as alert triage and prioritization, incident scoping and impact assessment, as well as IDS performance auditing. The contributions of this work are summarized as follows:

\begin{itemize}
\item The development of a high-fidelity framework to enhance cybersecurity operations in connected-vehicle infrastructure networks.
\item The development of a reliable ground-truth labeling model for network traffic flows to improve the quality of training data for deployed IDSs.
\item A framework for post-flow network analysis to support SOC activities, including forensic analysis, threat hunting, and IDS auditing.
\item The determination of an upper bound on the performance of a system IDS under conditions of full visibility and information availability.
\end{itemize}

The remainder of this paper is structured as follows: Section \ref{rw} presents and analyzes the state of the art. Section \ref{sm} overviews the system model and contextualizes the integration of the FALCON-C framework within the broader EV charging ecosystem and cybersecurity operations. Section \ref{meth} outlines the methodology used in this work. Section \ref{ra} presents and analyzes the experimental results. Finally, Section \ref{conc} concludes the paper and discusses opportunities for future work.

\section{Related Work}
\label{rw}

This section outlines the state of the art regarding network intrusion detection for EVSE infrastructure. The three prevalent approaches in the literature for ML-based implementations are \textbf{(A)} Neural Network Models, \textbf{(B)} Traditional ML Models, and \textbf{(C)} Anomaly Detection Models. It should be noted that the first two categories model the task as classification rather than as an anomaly-detection task; hence, the separation and creation of a third category exclusively focusing on works that frame the problem as anomaly detection. The selected works focus either on the CICEVSE2024 dataset or, more broadly, on OCPP attack detection.

\subsection{Neural Networks}
Neural networks overwhelmingly comprise the majority of the available literature. Some of the seminal works published, Morosan and Pop \cite{Morosan2017}, and Kabir \textit{et al.} \cite{Kabir2021}, describe backpropagation neural networks to monitor OCPP traffic. Morosan and Pop do not train their model on OCPP attacks; instead, they attempt to classify traffic as either normal or faulted. They also attempt to distinguish between normal and randomly generated non-faulted traffic. Kabir \textit{et al.} adopt a different approach, training their model on time-based features extracted from the OCPP traffic with the goal of detecting a specific type of switching attack. 

These studies achieved a high accuracy in distinguishing between malicious and benign traffic, demonstrating early success in the field. Morosan and Pop report approximately 91\% accuracy, while Kabir \textit{et al.} report 92\% accuracy for attacks lasting 20 seconds, with accuracy reaching nearly 100\% after 30 seconds. However, they also report a high false negative rate, starting around 30\% at 20 seconds and only dropping to 10\% at 30 seconds. 

The works using the newer CICEVSE2024 dataset utilize various neural network models, with Long Short-Term Memory (LSTM) being the most common. Deep Neural Networks (DNNs) are the next most common, followed by Recurrent Neural Networks (RNNs) and Convolutional Neural Networks (CNNs). In general, these studies either attempt to apply binary labels (benign or malicious) or group attacks into specific categories (benign, DoS, or recon). The best results were obtained by Thapa \textit{et al.} \cite{Thapa2025} with an LSTM model achieving 99\% accuracy, recall, and precision for binary classification. Purohit and Govindarasu \cite{Purohit2024} obtain approximately a 97\% accuracy and F1 score with a DNN. Rahman \textit{et al.} \cite{Rahman2025} train a DNN, CNN, LSTM, and CNN-LSTM hybrid model to classify traffic as benign, recon, or DoS. The best performance was achieved by the CNN-LSTM model with an accuracy, recall, and precision each approximately 97\%.

\subsection{Traditional ML}
Few publications have approached the CICEVSE2024 dataset using non-neural network models. One of these is the original dataset paper by Buedi \textit{et al.} \cite{Buedi2024}, which reports modest detection results. The authors train various models, including decision trees, k-Nearest Neighbor, Adaboost, Multi-Layer Perceptron, Naïve Bayes, Logistic Regression, Random Forest, and Support Vector Machine on their dataset, achieving a maximum 94\% accuracy, 91.5\% precision, and 94\% recall for multiclass classification. Random Forest exhibits consistent accuracy, while also achieving the highest precision and F1 score at nearly 97\% and 95\%, respectively. 

Makhmudov \textit{et al.} \cite{Makhmudov2025} train an Adaptive Random Forest model that achieves 99\% accuracy, 99.9\% precision, and 99\% recall for binary classification, and 98\% accuracy, precision, and recall on multiclass classification. These results are very promising, as they outperform many more complex neural network approaches.

\subsection{Anomaly Detection}
The goal of anomaly detection models is to create a system capable of detecting previously unseen attack vectors. This is accomplished by modeling the standard behavior of benign traffic. Both works addressed in this section employ variations of autoencoder models to accomplish this, with mixed results. Jahangir \textit{et al.} \cite{Jahangir2024} and Terruggia \textit{et al.} \cite{Terruggia2025} both train autoencoder variants on features extracted from OCPP traffic. Jahangir \textit{et al.} report 91\% accuracy, 88\% recall, and 91\% recall when presented with a variety of price manipulation attacks. Terruggia \textit{et al.} report 97\% accuracy and 98.6\% recall, but only 52\% precision on their model trained to detect DDoS attacks. The majority of false positives appear to be clustered near times of actual attacks.

% The third publication, Kesavan et al., describes a complex intrusion detection system involving both an AE anomaly detection component and a supervised learning attack classification component \cite{Kesavan2025}. The dataset cited is no longer available on Kaggle but supposedly contained real-world data from EV charging sessions. Using this dataset as a baseline, they applied several attacks ranging from False Data Injection to DoS. The overall model achieved 96.8\% accuracy with only a 1.8\% false positive rate, but there were no specific numbers for only the AE portion of the model. 

\subsection{Literature Gap Addressed}
To the best of our knowledge, no work has attempted to perform true anomaly detection on benign traffic from CICEVSE2024, as some neural network-based works claim to perform anomaly detection while actually performing binary classification. Additionally, publications that do perform true anomaly detection on other datasets have room for performance improvement. This work will address this gap by introducing FALCON-C, an autoencoder-based flow-labeling and anomaly-detection module designed to improve the quality and availability of IDS training datasets and enhance SOC operations, including threat hunting/modeling, IDS output validation, and post-incident SOC analysis.

\section{System Model}
\label{sm}

\begin{figure*}[!hbt]
\centerline{\includegraphics[width=1.3\columnwidth]{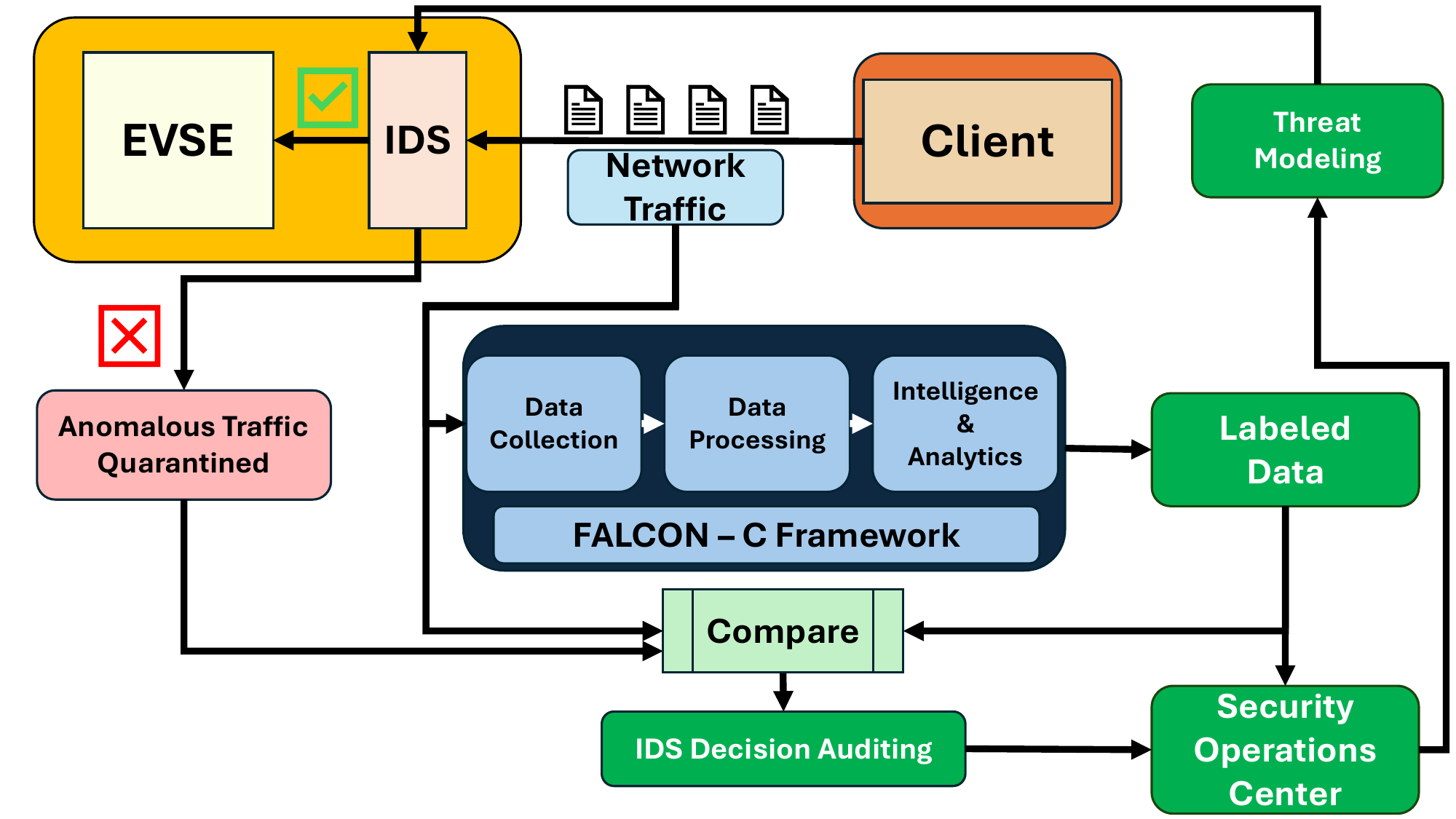}}
\caption{System Model with FALCON-C Framework Integration}
\label{fig:system_model}
\end{figure*}
The proposed system model is presented in Fig. \ref{fig:system_model}. At a foundational level, it depicts the interaction between a vehicular client and the EV charging infrastructure. Overlaying that foundation are the system's networking and cybersecurity components. As shown, the vehicular client generates network traffic as it communicates with the charging infrastructure to coordinate charging activities. This traffic passes through a deployed IDS that uses anomaly detection to distinguish between normal and anomalous traffic. The IDS aims to forward benign traffic to the EVSE or quarantine malicious traffic to prevent further harm to the system. Once traffic is received and identified, it is passed to the FALCON-C Framework, which collects network data, processes it into complete flows, and labels the flows as benign or malicious. 

Once labeled, this data is used for several tasks, the first of which is IDS decision auditing. By comparing the IDS's decision, made with partial flow visibility, to the FALCON-C label, derived from full flow statistics, the upper bound performance of the IDS can be determined and its performance assessed. This assessment can then be sent to the SOC for continuous model improvement. The labeled data is also sent directly to the SOC to support critical tasks, including automated threat hunting and modeling, alert triage, and incident scoping. The proposed system model demonstrates the integral role of the FALCON-C framework across various aspects of the system’s cybersecurity, including its elements, entities, and components.

\section{Methodology}
\label{meth}
This section describes the methodology of the FALCON-C Framework, including a comprehensive description of the dataset used, the preprocessing steps employed, and the architectural decisions surrounding the development and training of the autoencoder model.

\subsection{Dataset Description}
The CICEVSE2024 dataset is collected on two separately configured testbeds labeled EVSE-A and EVSE-B. EVSE-A focuses only on OCPP traffic. It is built with a Grizzl-E charger that reaches out to a remote CSMS. EVSE-B contains both OCPP and ISO 15118 traffic, as well as logs of kernel events and power consumption. It is built as a system of Raspberry Pis with the CSMS located on the local network. The attacker is a local network device in both testbeds \cite{Buedi2024}. Given the layout of typical EVSE infrastructure, it makes sense to place an IDS at the CSMS within the network topology to monitor all OCPP traffic flowing in both directions.

As OCPP 1.6 has implemented TLS, external detection systems must be built to detect attacks based on network and transport layer data, not application data. It should be noted that the CICEVSE2024 dataset contains only attacks that generate a high volume of traffic, meaning that these features are sufficient to detect abnormal activity. It is very likely that the detection of more sophisticated attacks involving OCPP application data manipulation would require inspection of the application data.

The EVSE-A and EVSE-B data are organized into separate files, each file containing a unique attack type and charging state. Each attack is seen during charging and idle sessions. The list of attacks is as follows:
\begin{itemize}
    \item Recon:
    \begin{itemize}
        \item TCP Port Scan, Service Version Detection, OS Fingerprinting, Aggressive Scan, SYN Stealth Scan, Vulnerability Scan
    \end{itemize}
    \item DoS:
    \begin{itemize}
        \item UDP Flood, ICMP Flood, PSHACK Flood, ICMP Fragmentation, TCP Flood, SYN Flood, SynonymousIP Flood, Slowloris Scan
    \end{itemize}
\end{itemize}

EVSE-A contains 10,078 benign charging packets and 64,061 benign idle packets. When converted to network flows, this results in 14 charging flows and 68 idle flows. In total, EVSE-A contains 547,854 network flows. EVSE-B contains 2,196,846 network flows with no benign traffic.

\subsection{Data Analysis and Preprocessing}
The network flows are created by processing the packet capture files with the NFStream Python framework. Each flow defines the series of packets sharing the same source and destination IP and port. The flows result in 86 features containing information such as start and end time, packet size range, number of packets, TCP flags, and application type.

The CICEVSE2024 dataset is split into many CSV files, each containing a specific attack type and charging state. These files are loaded into a DataFrame and labeled benign, recon, or DoS. During exploratory data analysis, several trends were uncovered. First, the dataset contains 86 features, several of which need to be encoded. After encoding IP addresses, a few source IP addresses are found to strongly correlate with benign traffic. By contrast, destination IP addresses do not correlate as strongly with any specific label. Features detailing packet flags were also analyzed, and no strong patterns were identified. A high number of ACK flags shows a slight correlation with benign traffic, which could reflect that benign traffic includes more successful connections than malicious traffic.

It should be noted that after assigning labels to the data, some columns were dropped because they were unsuitable or not useful for the present task. These included the ID column, MAC address OUIs, and some naïve guesses about the application associated with the traffic.

% \subsection{Random Forest}
% The Random Forest model was built with both EVSE-A and EVSE-B using scikit-learn. The data divided using an 70-30 train test split and then standardized. No model optimization was performed.

% \subsection{K Means Clustering}
% The K Means Clustering algorithm was largely a failure. It was built with the EVSE-A dataset and scikit-learn. No quantifiable patterns emerged from the generated clusters. With six clusters total, benign traffic was only found within two clusters, but there was far more malicious traffic under both of those labels. 

% \subsection{RNN}
% The RNN model was built with the EVSE-A and EVSE-B models using Tensorflow. The data was standardized with scikit-learn and a 80-20 train test split was applied. The RNN model is a shallow RNN trained to classify benign, recon, or DoS traffic for EVSE-A and recon or DoS for EVSE-B. The model suffers from underrepresentation of the benign traffic class, leading to a 0\% True Negative rate for the EVSE-A model. (This will be further analyzed in the Results and Analysis section).

\subsection{Autoencoder Architecture}
\begin{figure}[!hbt]
\centerline{\includegraphics[width=0.9\columnwidth]{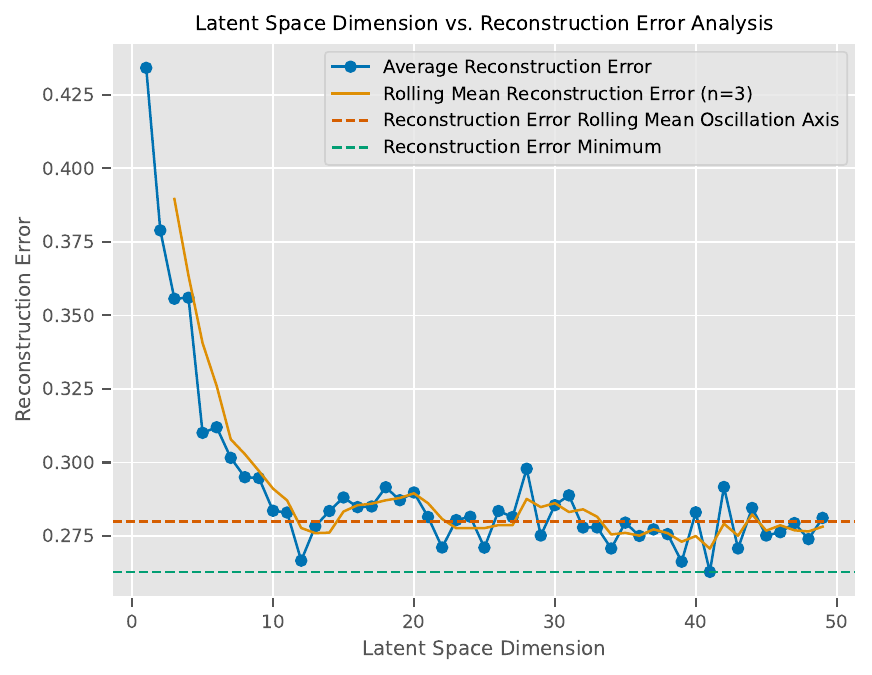}}
\caption{FALCON-C Autoencoder Latent Space Analysis}
\label{fig:latent_space_analysis}
\end{figure}

Regarding the autoencoder, only the EVSE-A dataset is used for training. The goal is to model the baseline behavior of normal traffic, in the hopes of obtaining a significantly different statistical reconstruction error profile for the malicious traffic. The data was standardized, and an 80-20 train-test split was applied. The autoencoder is built with TensorFlow and comprises one hidden layer in the encoder and decoder, with an additional hidden layer for the latent space representation of the data. The autoencoder had 151 input features after pre-processing, and the encoder/decoder hidden layer had 80 neurons, corresponding to approximately 50\% of the input feature space. Regarding the latent space, an experimental analysis was conducted to identify the optimal dimension. For this experiment, latent space dimensions along the range [1, 49] were assessed. Each value was used to train 5 models, with summary statistics of the results presented in Fig. \ref{fig:latent_space_analysis}.

As seen in Fig. \ref{fig:latent_space_analysis}, as the latent space dimension increases, the reconstruction error tends to decrease. This is especially prevalent for smaller values and becomes less impactful as the latent dimension increases beyond 10. After constructing a rolling average of the mean reconstruction error, it can be seen that the reconstruction error oscillates about a mean (denoted by the red horizontal line) in the figure. This suggests that there is no significant performance gain past a latent dimension of 10. However, the small number of individual trials for each latent space dimension is likely contributing to these results. It is hypothesized that a larger number of trials would reduce the oscillation about the axis, and the performance would achieve a steady state. In the interest of completeness, a latent space dimension of 41 was selected and implemented for the remainder of this work to correspond with the absolute minimum average reconstruction error exhibited across all latent space dimensions. Table \ref{tab:auto} lists relevant hyperparameters for the autoencoder.

\begin{table}[!hbt]
\caption{Autoencoder Hyperparameters}
\centering
\begin{tabular}{|c|c|}
\hline
\textbf{Hyperparameter}          & \textbf{Value} \\ \hline
Activation Function (all layers) & ReLU           \\ \hline
Optimizer                        & ADAM           \\ \hline
Loss                             & MSE            \\ \hline
Epochs                           & 100            \\ \hline
\end{tabular}
\label{tab:auto}
\end{table}

\section{Results and Analysis}
\label{ra}
This section outlines the various results obtained from the study, including the autoencoder training process, the autoencoder reconstruction error profile, and the anomaly detection decision boundary.
\subsection{Autoencoder Training Process}
\begin{figure}[!hbt]
\centerline{\includegraphics[width=0.9\columnwidth]{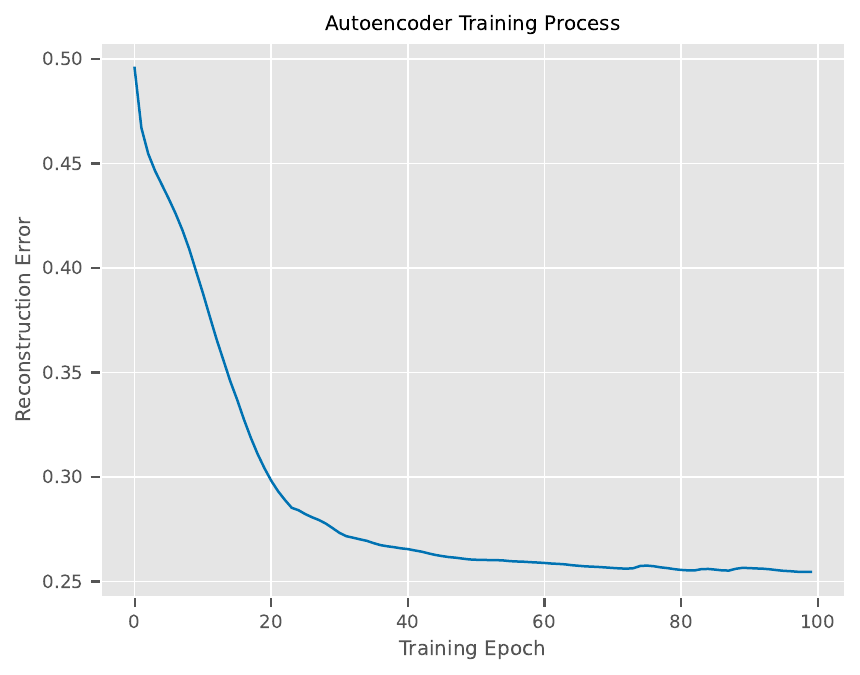}}
\caption{FALCON-C Autoencoder Training Process}
\label{fig:ae_training}
\end{figure}

The first result in Fig. \ref{fig:ae_training} shows the autoencoder’s training process, with the reconstruction error modeled as a function of the training epoch. As shown in this figure, the model exhibits a consistent training process, as evidenced by the smooth curve with a steady decline in reconstruction error over time. The model has converged to an acceptable reconstruction error, with early stopping triggered around the 100th training epoch. These initial insights into the model’s training process are promising, as a low reconstruction error has been achieved, suggesting that the selected autoencoder architecture is capable of effectively learning the latent representation of the network flows. The next analysis will examine the reconstruction error observed when faced with unseen flows and when malicious flows are introduced.

\subsection{Autoencoder Data Reconstruction Error Profile}

\begin{figure}[!hbt]
\centerline{\includegraphics[width=\columnwidth]{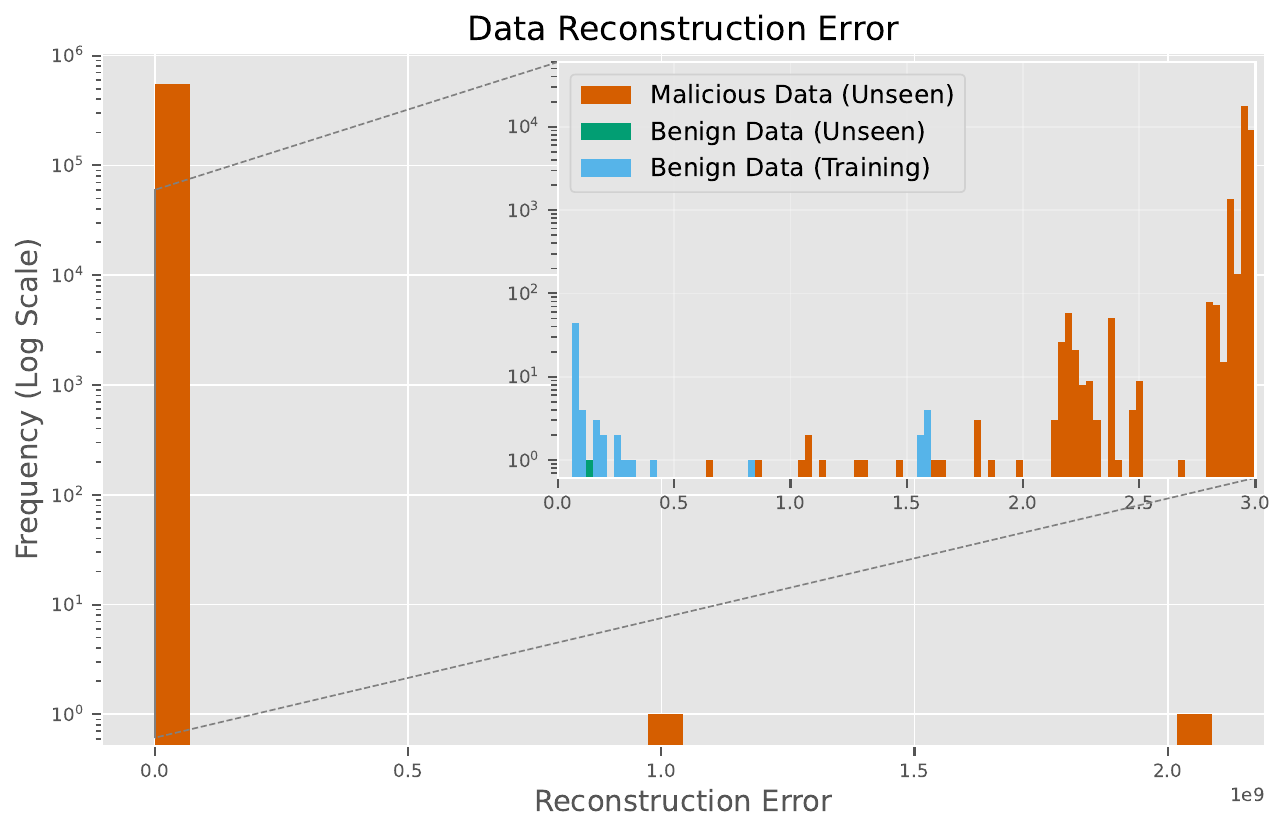}}
\caption{FALCON-C Autoencoder Data Reconstruction Error}
\label{fig:ae_reconstruction_error}
\end{figure}

The second set of results, presented in Fig. \ref{fig:ae_reconstruction_error}, shows the per-sample reconstruction error for the seen benign training data, unseen benign test data, and unseen malicious test data. Note that the y-axis of both the main and nested graphs uses a logarithmic scale. The first observation is the significant imbalance between malicious and benign flows. As previously mentioned, due to this imbalance, one of the main objectives of this work is to distinguish between benign and anomalous flows using a small set of benign flows for training. 

In the base figure, the reconstruction errors for the malicious flows form three main clusters, with the first cluster containing the vast majority of flows. This result is promising, as the remaining two clusters exhibit significantly higher reconstruction errors and will be easily distinguishable from benign flows. Unfortunately, the base figure's visualization is limited in providing insight into the separation of benign from anomalous flows within the first cluster, due to the x-axis scale. To address this, a zoomed-in portion of the figure is shown in the upper right corner, with the x-axis limited to reconstruction errors between 0 and 3 units. This region was selected because it includes all benign flows and can be used to determine appropriate decision thresholds for implementing the labeling. 

In this zoomed-in portion, a few key insights emerge. First, the majority of benign flows exhibit a reconstruction error of less than 0.5 across both seen and unseen data, and all unseen data (albeit limited in number) fall within this range. The critical observation in the range [0, 0.5] is the absence of any anomalous samples. This suggests that we can say with high confidence that a flow with a reconstruction error below 0.5 is benign. This result indicates an initial success for the FALCON-C autoencoder model in distinguishing between benign and anomalous traffic. Beyond the 0.5 error mark, a small group of benign training data shows a higher reconstruction error of approximately 1.6. There is also a single sample with a reconstruction error of approximately 0.7, located beside a malicious sample, suggesting a similar error profile. Both cases must be considered and further explored when determining an optimal decision boundary.

\subsection{Anomaly Detection Threshold Analysis}
\begin{figure*}[!hbt]
\centerline{\includegraphics[width=2\columnwidth]{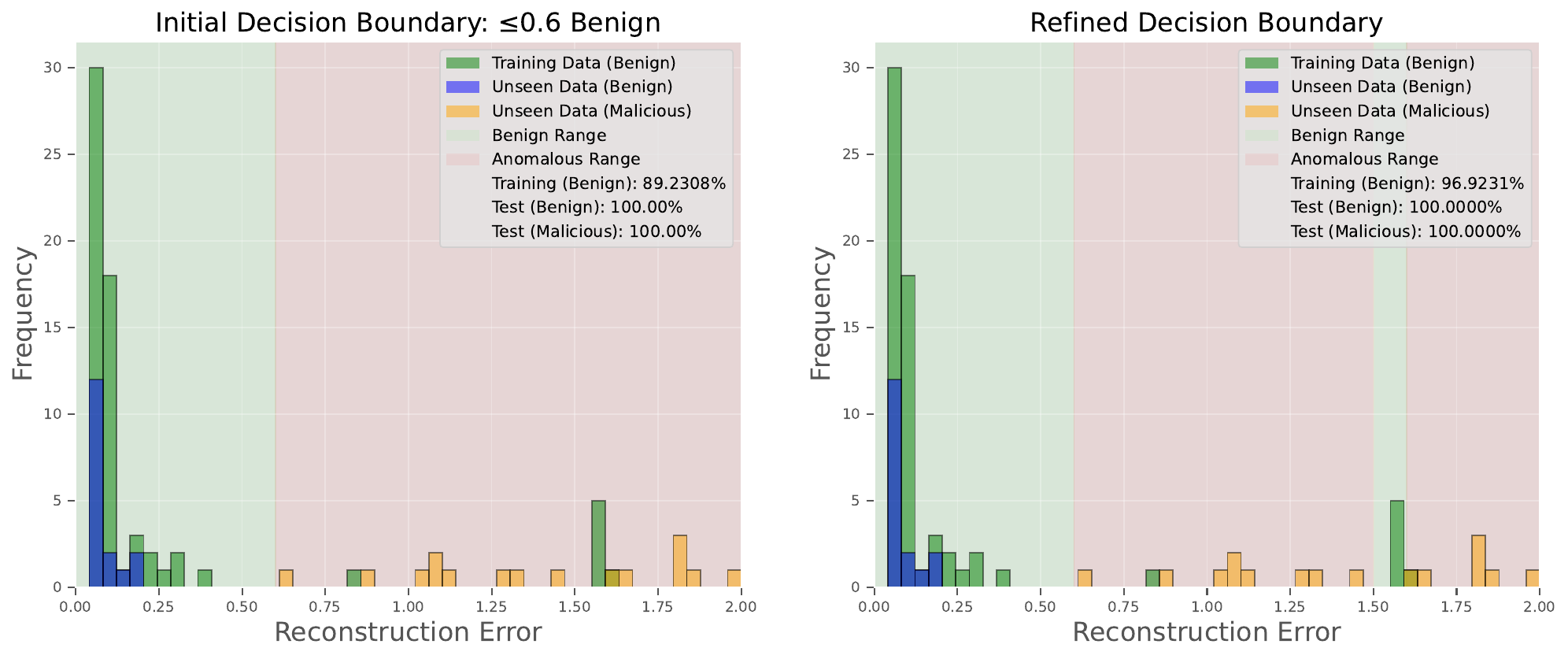}}
\caption{FALCON-C Autoencoder Decision Boundary}
\label{fig:ae_decision_boundary}
\end{figure*}

The third set of results, shown in Fig. \ref{fig:ae_decision_boundary}, presents two decision boundaries developed to classify flows as benign or malicious. Both figures highlight the decision boundaries with shaded green and red regions. Additionally, the distribution of seen and unseen data, separated by type, is overlaid on the decision boundary region to visualize the anomaly detection performance. Furthermore, performance statistics are listed in the legend for each decision boundary strategy.

The first decision boundary, shown on the left, represents an initial naïve thresholding strategy in which all flows with a reconstruction error less than or equal to 0.6 are classified as benign, and any flow exceeding 0.6 is deemed malicious. In the previous analysis, a value of 0.5 was discussed as the boundary; however, given the limited benign flow data, a value of 0.6 was selected to serve as a buffer in case new benign data with a slightly higher reconstruction error is observed. As shown in the figure, this thresholding approach achieved a training accuracy of 89.23\% and exhibited perfect labeling for all unseen samples. This result is promising, as the FALCON-C model achieved perfect anomaly detection while using only a select few benign flow samples during training. The results for the unseen benign flows are consistent, since the samples fall within the group of the majority of benign sample reconstruction error values. If a benign sample exhibited a reconstruction error profile similar to the training examples that are being mislabeled, it would also be mislabeled. This observation prompted the refinement of the decision boundary, as shown in the right figure.

The refined decision boundary leverages training performance to improve anomaly detection by reducing the false-positive rate. After training the FALCON-C autoencoder, the reconstruction errors showed a small number of samples at approximately 1.6. Intuitively, this is not a random observation but rather a symptom of the few benign training flows, and future flows are very likely to exhibit a similar reconstruction error profile. To this end, the decision boundary was partitioned to include the small range [1.5, 1.6] within the benign region. This minor adjustment improved the training accuracy by 8.62\%, yielding a training accuracy of 96.92\%. This result is significant because the detection and correct labeling of malicious flows were not affected and maintained perfect performance. 

\section{Conclusions and Future Work}
\label{conc}

The work presented in this paper proposes the \textbf{F}low-based \textbf{A}nalysis and \textbf{L}abeling for \textbf{CO}nnected vehicular \textbf{N}etwork \textbf{C}ybersecurity (FALCON-C) framework, an autoencoder-based mechanism for automatic network flow labeling in connected vehicular networks, with a direct application to Electric Vehicle Charging Infrastructure. The autoencoder architecture was determined through experimentation on the relationship between latent space dimension and reconstruction error. The FALCON-C framework demonstrated a smooth training process and achieved 100\% detection of malicious flows when trained exclusively on a small number of benign flows. Initially, using a naïve threshold scheme, the model exhibited a suboptimal false-positive rate, labeling a select number of benign outlier flows as malicious. This was remedied by studying FALCON-C's performance on the training data and using it to inform a more refined decision boundary. The result was a greatly reduced false-positive rate with no impact on the framework’s malicious flow detection capability. In summary, the FALCON-C framework has significant implications for cybersecurity operations in connected vehicular networks by enabling automated flow labeling and enhancing Security Operations Center activities through improved threat hunting, defense auditing, and incident scoping.

 Future work will continue refining the results of this study and exploring avenues for improvement. Specifically, a deeper understanding of the false-positive flows will be required. This will be accomplished by thoroughly examining the communication patterns that led to the mislabeled flows. These insights will then be used to engineer additional defining features that further refine the decision boundary and improve overall framework performance. Additionally, different autoencoder architectures will be explored, and an ensemble model will be developed to help address the mislabeled samples.

% The growing attack surface of EVSE infrastructure has led to the continued development of various datasets and machine learning models to support attack detection efforts in the field. This paper has shown improvements on previous works' results with the Random Forest model, reaching nearly 100\% accuracy. It has also contributed a new approach within the field, proposing the first true anomaly detection model built from the CICEVSE2024 dataset. These results demonstrate that these Denial of Service and Reconnaissance attacks are trivial to detect, and building Intrusion Detection Systems for EVSE infrastructure is a worthwhile endeavor.

% For future improvements, the RNN and Autoencoder models need more benign traffic to fully learn to identify that class. This would address the class imbalance issue with the RNN, and the autoencoder would have a more robust baseline and test set. Additionally, implementation of these models in a fully functioning IDS on a live testbed would further demonstrate their utility in the field of EVSE security. 

\bibliographystyle{IEEEtran}
\bibliography{sample}

% that's all folks
\end{document}